\documentclass[preprint]{aastex}

\slugcomment{To appear in ApJ}

\shorttitle{SNR in Turbulent Environment}
\shortauthors{Balsara, Benjamin, \& Cox}

\begin{document}

\newlength{\bigpicsize}
\setlength{\bigpicsize}{4.5in}
\newlength{\smpicsize}
\setlength{\smpicsize}{3.5in}


\title{The Evolution of Adiabatic Supernova Remnants in a Turbulent,
Magnetized Medium}


\author{Dinshaw Balsara}
\affil{NCSA, University of Illinois at Urbana-Champaign, Urbana, IL 61820}

\author{Robert A. Benjamin, and Donald P. Cox}
\affil{Department of Physics, University of Wisconsin-Madison, Madison, WI
53706}

\begin{abstract}
We present the results of three dimensional calculations for the MHD
evolution of an adiabatic supernova remnant in both a uniform and
turbulent interstellar medium using the RIEMANN framework of Balsara. In
the uniform case, which contains an initially uniform magnetic field, the
density structure of the shell remains largely spherical, while the
magnetic pressure and synchrotron emissivity are enhanced along the plane
perpendicular to the field direction. This produces a bilateral or
barrel-type morphology in synchrotron emission for certain viewing
angles. We then consider a case with a turbulent external medium as in
Balsara \& Pouquet, characterized by $v_{A}(rms)/c_{s}=2$. Several
important changes are found. First, despite the presence of a uniform
field, the overall synchrotron emissivity becomes approximately
spherically symmetric, on the whole, but is extremely patchy and
time-variable, with flickering on the order of a few computational time
steps. This is reminiscent of the Cas A observations of Anderson \&
Rudnick, although that remnant has a much more complicated pre-supernova
evolution than what is considered here. We suggest that the time and
spatial variability of emission in early phase SNR evolution provides
information on the turbulent medium surrounding the remnant. The
interaction of the outwardly-propagating SNR shock with interstellar
turbulence is shown to amplify the turbulence in the post-shock region,
thereby having important consequences for relativistic particle
acceleration. The shock-turbulence interaction is also shown to be a
strong source of helicity-generation and, therefore, has important
consequences for magnetic field generation. We compare
our calculations to the Sedov-phase evolution, and discuss how the
emission characteristics of SNR may provide a diagnostic on the nature of
turbulence in the pre-supernova environment.

\end{abstract}



\section{Introduction}

The morphology and emission of supernova remnants (SNRs) is in large part
determined by the nature of the surrounding circumstellar and
interstellar medium. Young SNRs from massive stars interact with a
significant mass of former wind material expelled by the star,
sometimes knotty and irregular, sometimes with significant preshaping
by the complex wind history. Older SNRs in the halo of the Galaxy will
have low X-ray surface brightness (Shelton et al 1998). SNRs in dense
molecular clouds will eventually be elongated in the direction of the
density gradient (Cox et al 1999).  SNRs at the edges of clouds will
develop offsets between the morphological and kinematic centers
(Dohm-Palmer et al 1996). Barrel-shaped remnants may probe the global
magnetic field structure (Gaensler 1998), and so on.

There have been several calls by observers, see Dickel, van Breugel \&
Strom (1991), Reynolds \& Gilmore (1993), Reed et al (1995) and others,
to introduce a realistic representation of interstellar turbulence in SNR
simulations. This need has been mirrored in the work of theorists, see
for example Jun \& Jones (1999), who claim that without the inclusion of
turbulence several observed effects are difficult to match. In
particular, Jun \& Jones (1999) find their simulated synchrotron indices
to be too smooth and uniform and not in line with the patchy observations
of Anderson \& Rudnick (1996). The problem of including MHD turbulence
in such simulations is daunting because of the high demands on resolution
and dynamic range. It is now feasible to do these simulations.

We have carried out the first 3D MHD simulations of a supernova going off
in a turbulent, magnetized ISM. For comparison
we have also carried out a simulation of an identical SNR going off in a
uniform but magnetized, non-turbulent ISM. The model includes no
pre-supernova wind, no radiative cooling, no thermal conduction and no
activity of a central pulsar. It is thus most relevant to the history of
Type 1a remnants prior to shell formation, exclusive of the
characteristics of the hot interior which would be altered by thermal
conduction. More significantly, however, it enlarges our intuition with
regard to analytic models, giving us some new insights into the
significant effects that interaction with turbulence can have.

\section{The Calculations}

The simulations presented here are calculated using the RIEMANN framework
for parallel, self-adaptive computational astrophysics described in
Roe \& Balsara (1996), Balsara (1998a,b), Balsara (1999a,b,c,d,e),
Balsara \& Spicer (1999a,b), Balsara \& Shu (2000), Balsara (2001a,b)
and Balsara \& Norton (2001). This framework uses some of the most
accurate higher order Godunov methods for non-relativistic
and relativistic MHD and radiative transfer that have been
devised. The advantages of this method,
particularly in modeling turbulent flows, is described in the above
references. The RIEMANN framework has been applied and tested on numerous
computational astrophysics problems that require numerical 3D-MHD. It has
been used to study the inverse cascade of magnetic helicity (Balsara \&
Pouquet 1999), star formation studies including the effects of turbulence
(Balsara, Crutcher \& Pouquet 2001; Balsara, Ward-Thompson \& Crutcher
2001), numerical MHD dynamos (Balsara 2000), and AMR-MHD simulations
of SNRs and shock-cloud interaction (Balsara 2001c).

We present the results of two high-resolution runs, $(256)^{3}$ zones,
comparing the results of evolution in a medium with a uniform and
turbulent medium. In both calculations, the physical length of the
computational domain is $L=60$ pc. The results presented here were
obtained using a uniform mesh and do not utilize the adaptive mesh
capability of RIEMANN. In the case of the uniform medium, the initial
hydrogen particle density is $1~{\rm cm^{-3}}$; the initial
temperature is 8000 K ($c_{s}=9.8~{\rm km~s^{-1}}$), the initial
velocities are zero and there is a uniform magnetic field with
strength $3.57 \mu G$ parallel to the x-axis. The energy input of the
supernova is introduced by increasing the thermal energy of all the
zones within 2 pc of the center of the computation grid so that the
total thermal energy integrated over a spherical volume is $10^{51}$
ergs. The particle density in this volume is also increased to $4~{\rm
cm^{-3}}$ , corresponding to an ejecta mass of $5~M_{\sun}$. The
simulations are evolved to the point when the shock reaches the
computational boundary. Since there is no cooling or other physical
effects that set a physical scale, these parameters may be rescaled
without requiring additional simulations.

In the case of the turbulent ISM, the initial conditions are the same as
above. The uniform magnetic field is still retained, as described in the
previous paragraph. However, we introduce random
Gaussian fluctuations in
the velocity and magnetic fields, corresponding to $v_{A}(rms)/c_{s}=2$
[$B(rms)=8.94 \mu G$] and $v(rms)/c_{s}=2$. (See Balsara \& Pouquet 1999
for
more details). As a result of the set up we have a uniform mean field
as well as a random magnetic field superposed on it. The fluctuating
field component is $2.5$ times larger than the uniform field component as
a result of the trans-Alfvenic initial conditions.
These initial conditions are
consistent with the observations of Lee \& Jokipii (1976), Beck (1991),
and Beck et al. (1996),
who find that the Galactic magnetic field has a uniform component as
well as a fluctuating component with the two components being in rough
equipartition. The Galaxy's ISM has Mach numbers in the range of one to
two
and
our use of $v_{A}(rms)/c_{s}=2$ would be considered a rather
strong turbulence. We made this choice because we seek to study
the evolution of supernova remnants in strong interstellar turbulence
as a counterpoint to the quiescent medium in the previous simulation.
As the blast wave sweeps outward, there is some evolution
in the preshock medium, with growing density fluctuations and decaying
magnetic and velocity fluctuations. These fluctuations quickly establish
an energy power spectrum of $E(k) \propto k^{-5/3}$ in the surrounding
medium well before the blast wave propagates a large distance. Thus
almost all of the SNR evolution takes place in a fully developed
interstellar turbulence. In the present simulations it is not possible
to drive the turbulence in order to sustain it at its originally
imposed level. This is because such driving mechanisms entail
imparting momentum fluctuations to the ISM gas and would, therefore,
interfere
with the SNR shock propagation. However, the largest energy-bearing
length scales have an eddy turnover time of $2.35 \times 10^{5}$ years
which
ensures that there is always sufficient turbulent energy available at the
smaller length scales to ensure that the SNR shock interacts with
a realistic turbulence during the entire course of the simulation. The
rms magnetic field decays less than 15\% during our calculated evolution.

\section{Results}

Despite the presence of a magnetic field, the pressure jump associated
with the shock is sufficiently strong that the explosion evolution
agrees quite well with the analytical Sedov solution. We find that the
position of the shell is given by $R_{Sedov} \propto t^{-2/5}$.  The
density profile is found to be self-similar and is also in good
agreement with the Sedov solution. However, limited resolution
prevents us from getting the full factor of four density jump right at
the shock front. At early times, the postshock density is 2.6 times
greater than the preshock density, increasing with age to 3.3 as more
zones are overtaken by the outward-propagating shock wave.
The final timestep shown in Figure 1 at $t=57,700$
years, has the following Sedov characteristics, $R=25.5$ pc,
$v_{shock}=174~{\rm km~s^{-1}}$, $T_{2}=4.2 \times 10^{5}$ K and
$p_{2}=5.2 \times 10^{-10}~{\rm dynes~cm^{-2}}$. The calculated values
agree with these analytical estimates. This is just about the oldest
this remnant could get without radiative cooling playing a significant
role, although with the turbulence there may be some over-dense spots
in which cooling might not be negligible. We intend to incorporate
this effect in future work.

\begin{figure}[ht!]
\includegraphics[angle=90,totalheight=\bigpicsize]{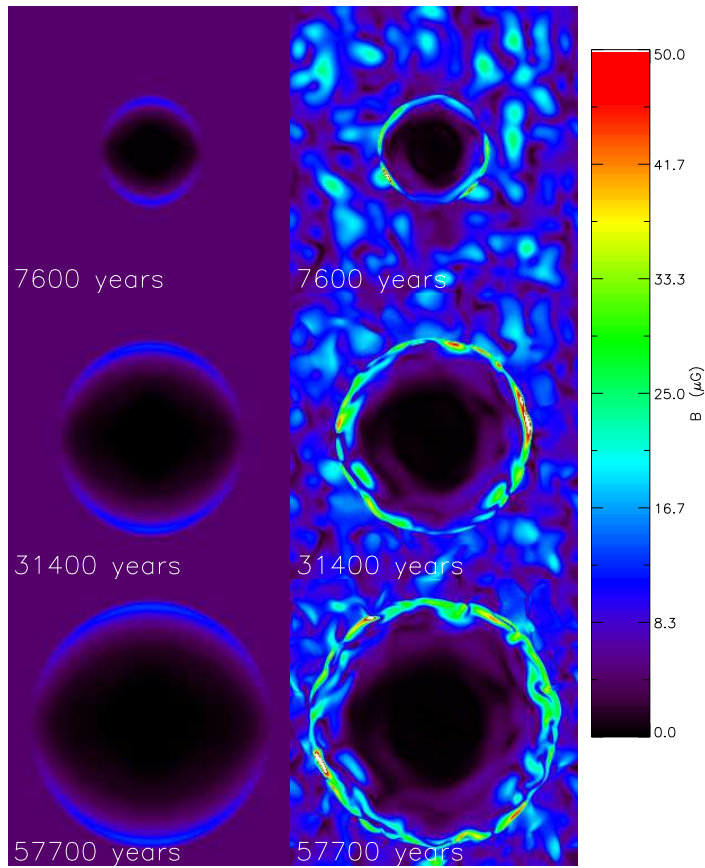}
\caption{{\small Total magnetic field strength for a two dimensional cut in
the XZ plane.  Panels on the left show magnetic field strength for the
case with a uniform interstellar medium, while the right panel shows
the evolution for a turbulent interstellar medium.} }
\label{fig1}
\end{figure}

The
most notable feature of the uniform ISM simulation is that the uniform
magnetic field, initially parallel to the x-axis, is swept up into a
compressed shell in the XZ plane (See Figure 1). Using an estimated
synchrotron emissivity of
$p B^{1.5}$ , this produces a bilateral (or barrel-like) morphology for
the SNR. Our assumption of a synchrotron emissivity that scales as $p
B^{1.5}$ is based on the work of Jun \& Jones (1999) who show that it is
a natural result of a momentum distribution of accelerated particles with
a power law index of 4.

\begin{figure}[ht!]
\includegraphics[angle=0,totalheight=\bigpicsize]{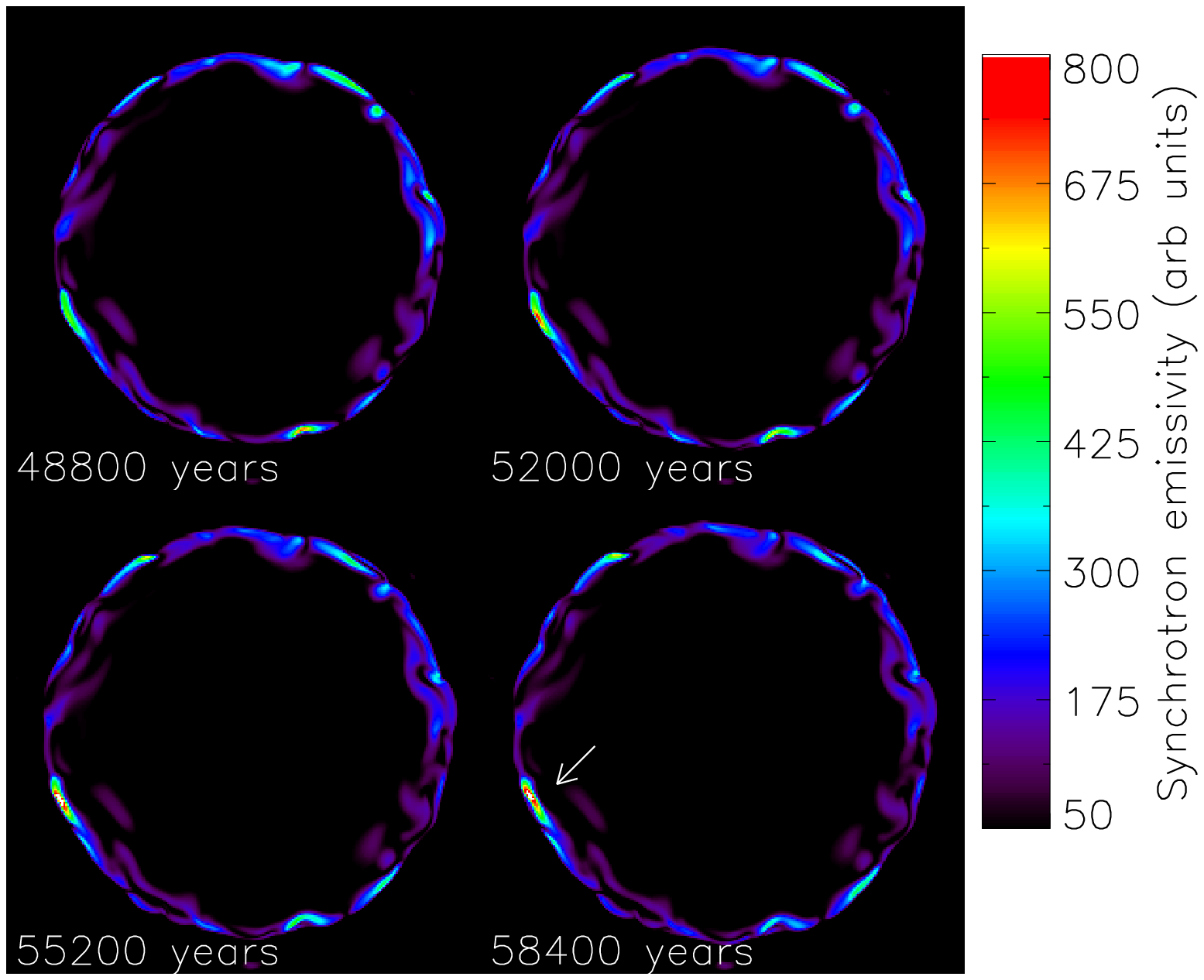}
\caption{{\small Estimated synchrotron intensity ($\epsilon \propto p
B^{1.5}$) for four closely related times. This shows large variablity
in intensity for different emission knots. Note, for example the
disappearance of bright spot in the lower right side of the remnant
and the appearance of bright knots on the upper and lower right sides.
This last emission knot is noted with an arrow.}}
\label{fig2}
\end{figure}

Several differences result for the case of a remnant evolving in a
turbulent medium. We have verified that the azimuthally averaged
density profile evolves self-similarly and agrees with the Sedov
solution.  However, in any given radial shell, there are significant
azimuthal variations in density just behind the blast wave. The ambient
medium is characterized by fluctuations which grow from $\delta \rho / \rho
\approx 0.1$ to 0.4 over the course of the simulation.  At the shock
front, the level
of density fluctuations, defined as $\frac{<\delta
\rho>}{\rho}_{post-shock}/
\frac{<\delta \rho>}{\rho}_{pre-shock}$, jumps by a factor of 4 at early
times.
At late times, this value declines
to $\sim 20$\% , presumably because
as the surface area of the remnant increases, individual lumps become
less important.  Behind the shock front, $\delta \rho / \rho$
decreases as the increased thermal pressure works to smooth out the
density fluctuations. This behavior is also consistent with one dimensional
shock-turbulence interaction studies reported in Balsara \& Shu (2000)
and has also been observed by several other authors cited in that work.

The variation in the magnetic field is shown in Figure 1 and is
especially noteworthy. The overall shape of the remnant is again
spherical, but with large
azimuthal variations. The mean magnetic field jumps by a factor of 2.4
across the shock front. This value is about what one would expect with
a factor of four jump in the two components of field perpendicular to
the shock front, i.e.
$B_{postshock}=\sqrt{(16B_{x,o}^{2}+16B_{y,o}^{2}+B_{z,o}^{2})/3}=3.3
B_{preshock}$.
The difference is due to the fact that resolution prevents us from getting
the full compressional effect. The random component of the
magnetic field in the ambient medium is  2.5 times larger than the uniform
component, so the barrel-like morphology noted in the uniform case is
absent. Kesteven \& Caswell (1987) have argued that a barrel morphology
would result from the circumstellar medium shaping the SNR.
Bisnovati-Kogan \& Silich (1995) and Gaensler (1998) considered
 the possibility that a uniform magnetic field
could give rise to a bilateral morphology. Our results suggest that in
order for an explanation that is based on magnetic fields to work,
the level of interstellar turbulence in the vicinity of the remnant
should be lower than the values we have considered. We have also carried
out a simulation with a turbulent Mach number of $1.5$ and we
found that it has the same trends as the Mach 2 simulation reported here.

A cross-section of the resulting synchrotron shell, shown in Figure 2,
is very patchy and
shows no evidence of the initial uniform magnetic field orientation.
While the magnetic pressure dominates
in the ambient medium, the pressure behind the blast wave is dominated
by thermal pressure for most of the evolution of the model
remnant. However, at late times, there are some spots in which the
magnetic pressure is approximately equal to the thermal
pressure. In these regions, the remnant can be synchrotron
bright and relatively X-ray faint.

These regions of unusually high field in the post-shock region tend to
be short lived in the absence of radiative cooling. This may be due in
part to dispersal by the enhanced thermal and magnetic pressures behind
the shock, but is almost certainly part of the natural evolution of the
turbulence. The result is visible in flickering in the brightest knots of
synchrotron emission, as in Figure 2, with flickering timescales that may
be related to knot intensity, a subject which we will be exploring in the
future. Such flickering, with 5-10 year timescales, has in fact been
reported for Cas A by Anderson \& Rudnick (1996), a remnant in which the
ejecta are apparently interacting with material from a presupernova wind.
Although this is very different from the model we have explored, the
overall structure is extremely knotty and it is possible that
irregularities (which must certainly be present in the magnetic field
within Cas A in order for the remnant to have such a knotty structure in the
synchrotron emission) behave similarly to those in our turbulent medium.
In our simulations, the timescale of this flickering is limited by
the lower size scale of the input spectrum of velocity and density
fluctuations, which in this case is about 10 grid cells. Higher resolution
simulations would yield much shorter timescale flickering.

\begin{figure}[ht!]
\includegraphics[angle=90,totalheight=\bigpicsize]{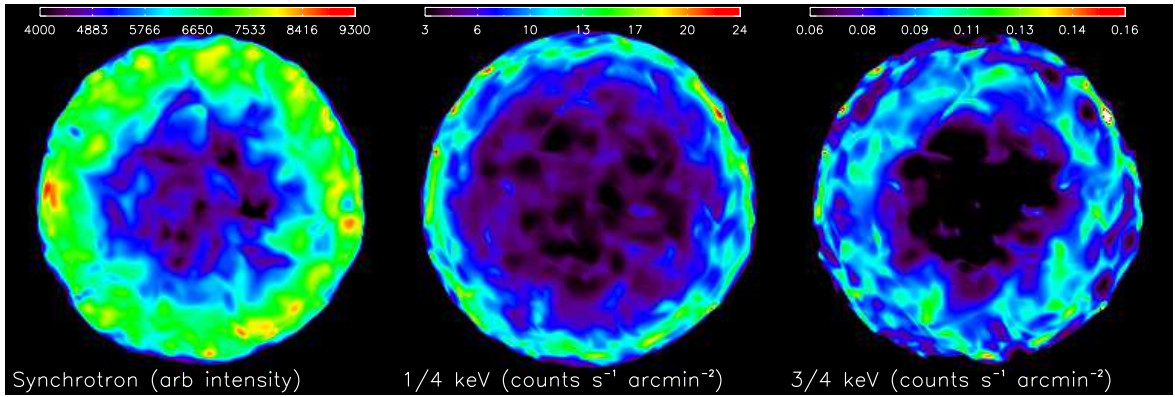}
\caption{{\small Three dimensional projection of emission onto the XY plane for (a)
ROSAT $\onequarter$ keV band, (b) ROSAT $\threequarters$ keV band, and (c)
synchrotron emission ($p B^{1.5}$). Note that the ROSAT $\onequarter$ keV
emission is more confined to a thin shell, while the ROSAT $\threequarters$
keV emission and synchrotron emission is more uniformly distributed.
There is some correlation between the synchrotron and X-ray emission.
However, there does not appear to be very strong point-to-point correlation
in any of the emission knots.}}
\label{fig3}
\end{figure}

We have also calculated the X-ray and estimated synchrotron
emissivities of the final time step for a three dimensional projection
onto each of the three principal planes of the turbulent
simulation. The X-ray emission is calculated assuming a plasma in
ionization equilibrium with a thermal spectrum as calculated by
Raymond \& Smith (1977) folded through the ROSAT response function for
the $\onequarter$ keV and $\threequarters$ keV emission bands. It is
assumed that all three bands are optically thin so that the total
emission may be obtained by summing the emissivities along each axis,
without any optical depth effects.  The results for the emission in
the XY plane are shown in Figure 3. We found that the fluctuations are
equally strong along all three projections both for the thermal and
non-thermal emission maps. As a result, the presence of the additional
uniform magnetic field leaves no imprint on the synchrotron
emissivity. The $\onequarter$ keV emission has a fluctuating but
fairly thin shell of emission, while the $\threequarters$ keV
emission, while fainter, is much more distributed in projected
radius. At this final timestep, the postshock temperature is
sufficiently low that the X-ray emission comes primarily from a region
somewhat back from the shock where the temperature is higher. This is
truer yet for the 3/4 keV band. Inclusion of thermal conduction may
fill the center further as in the study of W44 by Shelton et al
(1999).

\begin{figure}[ht!]
\includegraphics[angle=90,totalheight=\bigpicsize]{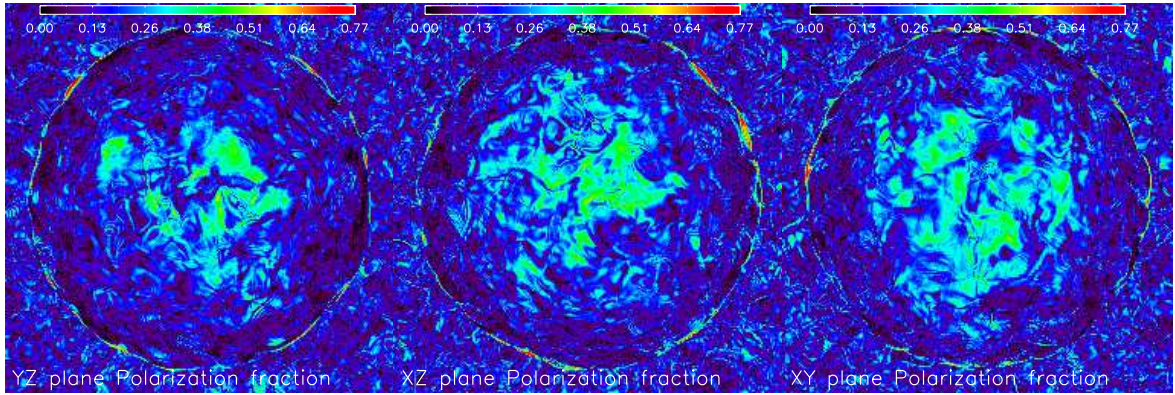}
\caption{{\small Synthetic synchrotron polarization maps showing the projections
along
the three principal axes at the same time step as Figure 3. The maximum
polarization of the synchrotron emission is 0.77, corresponding
to a electron energy power-law index of $p=2.5$. These images include
the effects of Faraday rotation in the remnant and external medium
for an observed frequency of 1 GHz. The turbulence considered here
is sufficiently strong that a signature of the initially uniform
field component does not show up in the polarization map at late times.}}
\label{fig4}
\end{figure}

We have also calculated the polarization of the synchrotron emission
as viewed along the three principal axes. We have assumed an electron
energy spectrum with power-law index $p=2.5$, which implies a maximum
linear polarization of 72\% (Longair 1994). Figure 4 shows the result
along all three axes for a frequency of $\nu=1$ GHz and includes
Faraday rotation arising in both the remnant and the
surrounding medium. We experimented with different frequencies and
different levels of Faraday rotation in the surrounding medium. In all
cases, we found that the level of turbulence we considered was large
enough the signature of the initially uniform field component does not
reveal itself in the polarization map that was made at a late time in
the simulation. Comparing Figures 3 and 4, it is very interesting to note 
that the brightest emission features
are also the most strongly polarized.

\begin{figure}[ht!]
\includegraphics[angle=0,totalheight=\smpicsize]{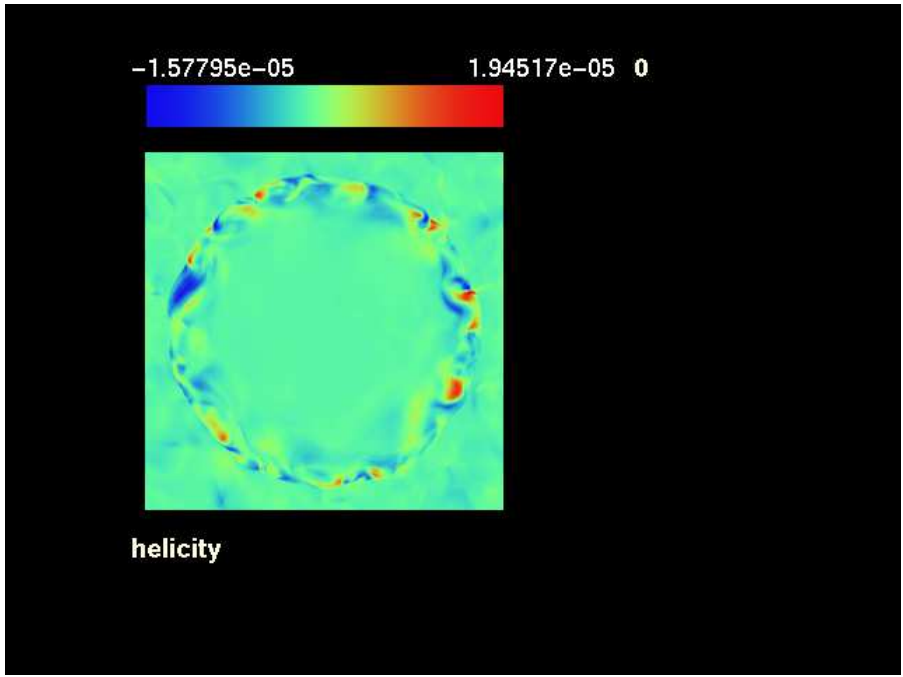}
\caption{{\small Color coded images for a two dimensional cut in the XZ
plane showing the fluid's helicity with both positive and negative values.
The units are ${\rm cm/sec^{2}}$.}}
\label{fig5}
\end{figure}

Several interesting points may be made about the interaction of
turbulence with strong shocks.
We focus here on one of the most interesting results,
the generation of helicity at shocks. Figure 5 shows the fluid
helicity in a cut in the XZ plane for the simulation where the SNR
evolves in a turbulent medium. The lower right panel in Figure 1 shows
the magnetic field strength in the same plane and at the same
time. The propagation of the SNR shock through the interstellar
turbulence has excited strong positive and negative helicity
fluctuations in the post-shock region. An examination of the magnetic
field strength shows that the field is strongest either in the regions
with the highest helicity or in the regions where helicities of
opposite signs closely abut one another. This positive correlation
between helicity and magnetic field is consistent with the stretch,
twist, fold scenario for magnetic field amplification that has been
detailed in Childress \& Gilbert (1995). Ferriere (1998) has presented
a scenario in which supernovae interacting with galactic
shear in a quiescent, stratified ISM could produce the helicity that
is needed in dynamo theories. The present simulations suggest that
interaction of strong supernova shocks with interstellar turbulence
might be an additional and highly dynamic source of helicity
generation and might, therefore, have relevance for fast dynamo
theories.

\section{Discussion}

Supernova remnants serve as useful ``test explosions'' that can be
used to probe the parameters characterizing the surrounding
circumstellar and interstellar medium.  These calculations inject an
element of realism into the modelling of the pre-SNR environment that
has heretofore been lacking, although much remains to be done. In a
quiescent medium, we find that the remnant sweeps up the uniform
field, and yields a bilateral morphology.  The introduction of
realistic, strongly turbulent fluctuations in the uniform medium,
however, erases this asymmetry, and produces remnants that are (on
average) spherical. The resulting projected X-ray and synchrotron maps
show a range of fluctuation in the emission characteristics of the
remnant.

We find that time variability and the angular variations in emission in
the shell are associated with the nature of the magnetic fluctuations in
the unshocked interstellar medium. We suggest, therefore, that it should
be possible to use the variation in emission in the remnant as a probe of
the turbulence in the pre-supernova environment.
A grid of models like these will be needed to determine how the nature of
the turbulence is related to the emission characteristics of the remnant.

Finally, we note that the interaction of SNR blastwave with a region of
supersonic MHD turbulence gives an opportunity to investigate the degree
to which magnetic fluctuations affect the effective equation of state of
magnetized plasmas. The importance of understanding the nature of this
behavior in interstellar conditions has been emphasized by McKee \& Zweibel
(1995). In future explorations, we will check to see how this
effective equation of state varies over time with the parameters
characterizing the MHD turbulence.

\acknowledgments This work was supported by NASA Astrophysical Theory
Grant NAG5-8417 to the University of Wisconsin-Madison. DB acknowledges
support from NSF grant 1-5-29014. We also thank NCSA
and SDSC for use of their supercomputers. We also wish to thank J.Dickel and
E. Zweibel for useful discussions.

\clearpage

%% No more than seven \figcaption commands are allowed per page,
%% so if you have more than seven captions, insert a \clearpage

%% There must be a \figcaption command for each legend. Key the text of the
%% legend and the optional \label in curly braces. If you wish, you may
%% include the name of the corresponding figure file in square brackets.
%% The label is for identification purposes only. It will not insert the
%% figures themselves into the document.
%% If you want to include your art in the paper, use \plotone.
%% Refer to the on-line documentation for details.

%% Tables should be submitted one per page, so put a \clearpage before

%% Two options are available to the author for producing tables:  the
%% deluxetable environment provided by the AASTeX package or the LaTeX
%% table environment.  Use of deluxetable is preferred.
%%

%% Three table samples follow, two marked up in the deluxetable environment,
%% one marked up as a LaTeX table.

%% In this first example, note that the \tabletypesize{}
%% command has been used to reduce the font size of the table.
%% Note also that the \label command needs to be placed

%% The following command ends your manuscript. LaTeX will ignore any text


\begin{thebibliography}{}

\bibitem[Anderson(1996)]{and96}Anderson, M.C. \&  Rudnick, L.,1996, ApJ,
456, 234

\bibitem[Balsara(1998a)]{bal98a}Balsara, D.S. 1998a, ApJS, 116, 119

\bibitem[Balsara(1998b)]{bal98b}Balsara, D.S. 1998b, ApJS, 116, 133

\bibitem[Balsara(1998a)]{bal99a}Balsara, D.S. 1999a, J. Quant. Spectros.
Rad.
Transf., 61, 617

\bibitem[]{}Balsara, D.S. 1999b, J. Quant. Spectros. Rad.
Transf., 61, 629

\bibitem[]{}Balsara, D.S. 1999c, J. Quant. Spectros. Rad.
Transf., 61, 637

\bibitem[]{}Balsara, D.S. 1999d, J. Quant. Spectros. Rad.
Transf., 62, 167

\bibitem[]{}Balsara, D.S. 1999e, J. Quant. Spectros. Rad.
Transf., 62, 225

\bibitem[]{}Balsara, D.S. and Spicer, D.S. 1999a, J. Comput. Phys., 148, 133

\bibitem[]{}Balsara, D.S. and Spicer, D.S. 1999b, J. Comput. Phys., 149, 270

\bibitem[]{}Balsara, D.S. and Pouquet, A. 1999, Phys. Plas., 6, 89

\bibitem[]{}Balsara, D.S. and Shu, C.-W. 2000, J. Comput. Phys., 160, 405

\bibitem[]{}Balsara, D.S. 2000, in  ``Astrophysical Plasmas: Codes, Models,
and Observations'', eds. S. J. Arthur, N. Brickhouse, and J. Franco,
RevMexAA(SC), 9, 92.

\bibitem[]{}Balsara, D.S., Crutcher, R.M. and Pouquet, A. 2001, ApJ, in
press

\bibitem[]{}Balsara, D.S. Ward-Thompson, D. and Crutcher, R.M. 2001, MNRAS,
in press

\bibitem[]{}Balsara, D.S. 2001a, ApJS, 132, 1

\bibitem[]{}Balsara, D.S. 2001b, J. Quant. Spectros. Rad.
Transf., 69(6), 671

\bibitem[]{}Balsara, D.S. 2001c, J. Comput. Phys.,"Divergence-Free Adaptive
Mesh Refinement for Magnetohydrodynamics", submitted

\bibitem[]{}Balsara, D.S. and Norton, C. 2001, Parallel Comput., 27, 37

\bibitem[]{}Beck, R., 1991 in ``ASP Conference Series Vol 18'', eds. N.Duric
and P.C.Crane.

\bibitem[]{}Beck, R.,Brandenburg,A., Moss,D.,Shukurov,A and Sokoloff,D.D.,
1996, ARAA, 34, 155

\bibitem[]{}Bisnovati-Kogan,G.S. and Silich,S.A., 1995, Rev. Modern Phys.,
67, 661

\bibitem[]{}Childress, S. and Gilbert, A.D. ,1995, "Stretch Twist Fold : The
Fast Dynamo", Springer Lecture Notes in Physics

\bibitem[]{}Cox, D. P., Shelton, R. L., Maciejewski, W., Smith, R. K.,
Plewa, T., Pawl, A., \& Rozyczka, M., 1999, ApJ, 524, 179

\bibitem[]{}Dickel, J.R., van Bruegel, W.R.J. \& Strom, R.G., 1991, AJ, 101,
2151

\bibitem[]{}Dohm-Palmer, R.C. and Jones, T.W., 1996, ApJ, 471, 279

\bibitem[]{}Ferriere, K. , 1998, ApJ, 335, 488

\bibitem[]{}Gaensler, B.M.,  1998, ApJ, 493, 781

\bibitem[]{}Jun, B.I. \& Jones, T.W., 1999, ApJ, 511, 774

\bibitem[]{}Kesteven, M.J. and Caswell, J.L., 1987, A\&A, 183,
118

\bibitem[]{}Lee, L.C. and Jokipii, J.R., 1976, ApJ, 206, 735

\bibitem[]{}Longair, M.S. 1994, High Energy Astrophysics: Volume 2,
(Cambridge: Cambridge University Press), 255

\bibitem[]{}McKee, C.F. \& Zweibel, E.G. 1995, ApJ, 40, 686

\bibitem[]{}Reed, J.E., Hester, J., Fabian,A.C. and Winkler, P.F.,1995, ApJ,
440, 706

\bibitem[]{}Reynolds, S.P. \&  Gilmore, D.M.,1993, AJ, 106,

\bibitem[]{}Roe,P.L. and Balsara, D.S. 1996, SIAM J. Appl. Math., 56, 57

\bibitem[]{}Shelton, R.L., 1999,  ApJ, 521, 217

\bibitem[]{}Shelton, R.L. et al. 1999,  ApJ, 524, 179

\bibitem[]{} Raymond, J.C. \& Smith, B. 1977, ApJS, 35, 419

\end{thebibliography}
\end{document}